\documentclass[prb,twocolumn,showpacs]{revtex4}

\usepackage{graphicx}
\usepackage{rotating}
\usepackage{amsmath}
\usepackage{amsfonts}
\usepackage{amssymb}
\usepackage{enumerate}
\usepackage{longtable}
\usepackage{dcolumn}
\usepackage{bm}

\begin{document}

\addtolength{\textheight}{8mm}

\draft

\title{\bf Raman Response of Magnetic Excitations in Cuprate Ladders and Planes}

\author{K.P. Schmidt$^{1,}$}
\email{ks@thp.uni-koeln.de}
\homepage{http://www.thp.uni-koeln.de/~ks}
\author{A. G\"ossling$^2$}
\affiliation{$^1$Institut f\"ur Theoretische Physik,
Universit\"at zu
  K\"oln, Z\"ulpicher Str. 77, D-50937 K\"oln, Germany\\
 $^2$II. Physikalisches Institut, Universit\"at zu
  K\"oln, Z\"ulpicher Str. 77, D-50937 K\"oln, Germany}
\author{U. Kuhlmann}
\author{C. Thomsen}
\affiliation{Institut f\"ur Festk\"orperphysik, Technische
Universit\"at Berlin, Hardenbergstr. 36, D-10623 Berlin, Germany}
\author{A. L\"offert}
\author{C. Gross}
\author{W. Assmus}
\affiliation{ Physikalisches Institut, J.W. Goethe-Universit\"at,
Robert-Mayer-Str. 2-4, D-60054 Frankfurt a.~M., Germany }
\date{\rm\today}

\begin{abstract}
An unified picture for the Raman response of magnetic excitations
in cuprate spin-ladder compounds is obtained by comparing
calculated two-triplon Raman line-shapes with those of the
prototypical compounds SrCu$_2$O$_3$ (Sr123),
Sr$_{14}$Cu$_{24}$O$_{41}$ (Sr14), and La$_6$Ca$_8$Cu$_{24}$O$_{41}$ (La6Ca8). The theoretical model for the
two-leg ladder contains Heisenberg exchange couplings $J_\parallel$ and
$J_\perp$ plus an additional
four-spin interaction $J_{\rm cyc}$. Within this model Sr123 and Sr14 can be
described by $x:=J_\parallel/J_\perp=1.5$, $x_{\rm cyc}:=J_{\rm
  cyc}/J_\perp=0.2$, $J_\perp ^{\rm Sr123}=1130$ cm$^{-1}$ and $J_\perp ^{\rm
  Sr14}=1080$ cm$^{-1}$. The couplings found for La6Ca8 are $x=1.2$, $x_{\rm cyc}=0.2$, and $J_\perp ^{\rm La6Ca8}=1130$ cm$^{-1}$.
The unexpected sharp two-triplon peak in the ladder materials
compared to the undoped two-dimensional cuprates can be traced
back to the anisotropy of the magnetic exchange in rung and leg
direction. With the results obtained for the isotropic ladder
we calculate the Raman line-shape of a two-dimensional square
lattice using a toy model consisting of a vertical and a horizontal ladder. A direct comparison of these results with Raman
experiments for the two-dimensional cuprates R$_2$CuO$_4$
(R=La,Nd), Sr$_2$CuO$_2$Cl$_2$, and YBa$_2$Cu$_3$O$_{6+\delta}$
yields a good agreement for the dominating
two-triplon peak. We conclude that short range quantum
fluctuations are dominating the magnetic Raman response in both, ladders and planes. We
discuss possible scenarios responsible for the high-energy
spectral weight of the Raman line-shape, i.e. phonons, the
triple-resonance and multi-particle contributions.
\end{abstract}
\pacs{75.40.Gb, 75.50.Ee, 75.10.Jm}
\maketitle


\section{Introduction}

\label{intro}

Strongly correlated electron systems in low dimensions are of
fundamental interest due to their fascinating properties resulting
from strong quantum fluctuations \cite{milli00,sachde00,ander00}. Especially
 in the case of the high-T$_{\rm c}$ cuprate
superconductors, the role of quantum fluctuations is heavily debated.
Two-magnon Raman scattering has been proven to be a powerful tool to study
quantum fluctuations in the magnetic sector\cite{sugai90,sulew90,knoll90,sulew91,yoshi92,blumb96}. In contrast to the well understood magnon dispersion as
measured by inelastic neutron
scattering\cite{manou91,kastn98,clark99,hayde91,colde01}, the quantitative
understanding of the two-magnon line-shape in the Raman
response\cite{sugai90,blumb96} and in the optical
conductivity\cite{perki93,perki98,gruni00,loren99} remains an issue open to debate.

Interestingly, in the so-called cuprate ladder systems like Sr123 or the telephone-number
compounds (Sr,Ca,La)$_{14}$Cu$_{24}$O$_{41}$ a prominent peak in the magnetic
Raman response is observed at the same energy of about $3000$ cm$^{-1}$ as in
the two-dimensional compounds\cite{goess03,sugai99,popov00,gozar01}.
In contrast to the gapless long-range ordered two-dimensional
compound, the quasi one-dimensional two-leg ladders are known to
be realizations of a gapped spin liquid\cite{dagot96}. Because the elementary
excitations above this groundstate are triplons\cite{schmi03c},
we call the corresponding Raman response as two-triplon Raman
scattering.

On the one hand, one may expect that the Raman response is dominated by
short-range, high-energy excitations, suggesting a certain similarity between ladders
and planes, both being built from edge-sharing Cu$_4$ plaquettes. The peak
frequencies are in fact at $3000$ cm$^{-1}$.
On the other hand, the line shape and in particular the peak width
strongly varies between different compounds. In 2D, the peak width is of the order of $1000$ cm$^{-1}$,
in La6Ca8 about $500$ cm$^{-1}$, in Sr123 and Sr14 only $(100-200)$ cm$^{-1}$.
Due to the observation of a very sharp two-triplon Raman line in the
spin liquid Sr14, Gozar {\it et al.} have questioned whether the large line width in 2D and the
related, heavily discussed spectral weight above the two-magnon peak can be attributed to
quantum fluctuations\cite{gozar01}.

In the last years theoretical developments in the field of quasi
one-dimensional systems, namely the quantitative calculation of
spectral densities\cite{knett01,schmi01,nunne02,gruni03,zheng03a,schmi04,trebs00}, has
led to a deeper understanding of magnetic contributions to the
Raman response of undoped cuprate ladders. Besides the usual Heisenberg exchange
terms the minimal magnetic model includes four-spin interactions which are 4-5
times smaller than the leading Heisenberg
couplings\cite{brehm99,nunne02,schmi03b,goess03,gruni03}. The existence and the
size of the four-spin interactions are consistent with
theoretical derivations of generalized $t$-$J$ models from
one-band or three-band Hubbard
models\cite{takah77,roger89,macdo90,schmi00,mulle02,mizun97,reisc04}.

In the present paper we show that the strong variation of
the line width can be traced back to changes of the spatial anisotropy of the exchange
constants. The sharp Raman line in Sr14 and Sr123 results from $x=1.5$, the
increased line width in La6Ca8 reflects $x=1.2$, and the isotropic couping
$x=1$ for the square lattice yields the much larger width observed in 2D. In
fact, we obtain a quantitative description of the dominant Raman peak in 2D
using a toy model which mimics the 2D square lattice by the superposition of a
vertical and a  horizontal ladder.
We thus conclude that the dominant Raman peak is well described by
short-range excitations.

Besides the dominant two-triplon peak, the large spectral weight
measured at high energies remains an open problem for the cuprate
ladders and planes. We review possible sources of
the high-energy spectral weight which were suggested in the past, e.g. quantum
fluctuations\cite{singh89,canal92,honda93,sandv98,katan03,schmi01}, the role of
spin-phonon interaction\cite{weber89,knoll90,saeng95,reill96,haas94,nori95,erole99} and
the triple resonance\cite{chubu95,schoe97,morr97,blumb96}. In case of the cuprate planes no final conclusion concerning the origin
of the high-energy weight can be drawn, but in the case of the cuprate ladders
the spin-phonon coupling and the triple resonance can be ruled out.

\section{Model}

\label{model}

In Raman scattering multi-particle excitations with zero change of the total
spin can be measured. Starting at $T=0$ from an $S=0$ ground state the
singlet excitations with combined zero momentum are probed. The Raman
response in spin ladders has been calculated by first order
perturbation theory\cite{Jurecka1} and by exact
diagonalization\cite{Suzuki1}. In this work, Raman
line-shapes are presented obtained from continuous unitary
transformations (CUT) using rung triplons as elementary
excitations\cite{schmi01,schmi03b}. The results are not resolution
limited because neither finite size effects occur nor an artificial
broadening is necessary.

\begin{figure}[htbp]
    \begin{center}
     \includegraphics[width=\columnwidth]{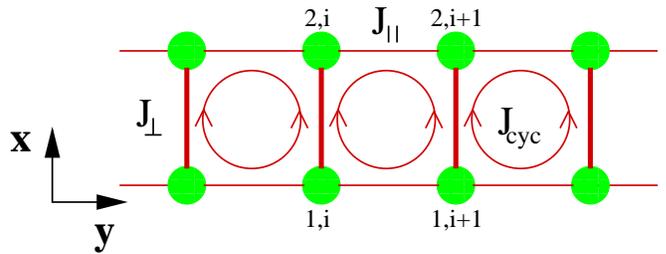}
    \end{center}
    \caption{Schematic view of a two-leg ladder (notation as in
      Eq. (\ref{eq:Hamiltonian})). The circles denote the positions of Cu$^{2+}$
       ions carrying a spin 1/2 each. The crystallographic axes
       are such that $x\parallel b$ and $y\parallel a$ for
       Sr123 and $x\parallel a$ and $y\parallel c$ for Sr14.}
    \label{fig_sketch}
\end{figure}

For zero hole doping, the minimum model for the magnetic
properties of the $S=1/2$ two-leg ladders is an
antiferromagnetic Heisenberg Hamiltonian plus a cyclic four-spin
exchange term $H_{\rm cyc}$ \cite{matsu00a,matsu00b,nunne02}
\begin{subequations}
\label{eq:Hamiltonian}
\begin{eqnarray}
\label{hamil1}
&&H = J_\perp \sum\limits_i {\bf S}_{1,i}
{\bf S}_{2,i} + J_\parallel \sum_{i,\tau} {\bf S}_{\tau,i} {\bf S}_{\tau,i+1}
+H_{\rm cyc}\\
\label{hamil2}
&&H_{\rm cyc}= J_{\rm cyc}\sum_i K_{(1,i),(2,i),(2,i+\!1),(1,i+\!1)}
 \\
&&K_{(1,1),(1,2),(2,2),(2,1)} = K_{1234} = \\\nonumber
&&\qquad({\bf S}_1 {\bf S}_2)({\bf S}_3 {\bf S}_4) + ({\bf
S}_1 {\bf S}_4)({\bf S}_2 {\bf S}_3) - ({\bf S}_1 {\bf S}_3)({\bf S}_2 {\bf
S}_4)\quad
\end{eqnarray}
\end{subequations}
where $i$ denotes the rungs and $\tau\in\{1,2\}$ the legs.
The exchange couplings along the rungs and along the legs are denoted by
$J_\perp$
and $J_\parallel$, respectively. The relevant couplings modeling
Sr123 and  Sr14\cite{sr14a} are illustrated in Fig.\ \ref{fig_sketch}.  There is also another way to include the
 leading four-spin exchange term by cyclic permutations \cite{brehm99,nunne02} which
differs in certain two-spin terms from Eq.\ (\ref{eq:Hamiltonian})
\cite{brehm99}.
Both Hamiltonians are identical except for couplings along the diagonals
if $J_\perp$ and $J_\parallel$ are suitably redefined \cite{notiz1}.

At $T=0$ the Raman response $I(\omega)$ is given by the retarded resolvent
\begin{equation}
 \label{Intensity}
 I(\omega) = -\frac{1}{\pi} {\rm Im}\left\langle0\left|
\mathcal{O}^{\dagger}(\omega-H+i\delta)^{-1}\mathcal{O}\right|0\right\rangle \ .
\end{equation}
The observables $\mathcal{O}^{\rm rung}$ ($\mathcal{O}^{\rm leg}$) for magnetic light
scattering in rung-rung (leg-leg) polarization read in leading order
\cite{Fleury1,Shastry1}
\begin{subequations}
\label{Observables}
\begin{eqnarray}
 \mathcal{O}^{\rm leg}& =& A_{0}^{\rm leg} \sum_{i}
\left( {\bf S}_{1,i}{\bf S}_{1,i+1} + {\bf S}_{2,i}{\bf S}_{2,i+1} \right) \\
 \mathcal{O}^{\rm rung}& =& A^{\rm rung}_{0} \sum_{i} {\bf S}_{1,i}{\bf S}_{2,i} \ .
\end{eqnarray}
\end{subequations}
The factors $A_{0}^{\rm leg}$ and $A^{\rm rung}_{0}$ depend on the
underlying
microscopic electronic model. It is beyond the scope of the present
work to compute them. The results will be given in units of these factors
squared. In this article we will only consider
non-resonant Raman excitation processes. We discuss which laser energy should be used in order
to investigate the non-resonant regime.

\section{Method}

\label{method}

Technically,  we employ a CUT to map the Hamiltonian $H$ to an
effective Hamiltonian $H_{\rm eff}$ which conserves the number of
rung-triplons, i.e.\ $[H_{\rm 0},H_{\rm eff}]=0$ where $H_{\rm
0}:=H|_{\{J_{\rm \parallel}=0;J_{\rm cyc}=0\}}$
\cite{Uhrig1,knett03a,knett04}. The ground state of $H_{\rm eff}$
is the rung-triplon vacuum. For the response function $I(\omega)$
the observable $\mathcal{O}$ is mapped to an effective observable
$\mathcal{O}_{\rm eff}$ by the same
CUT. The CUT is implemented in a perturbative fashion
in $x=J_{\parallel}/J_{\perp}$ and $x_{\rm cyc}=J_{\rm
cyc}/J_{\perp}$. The effective Hamiltonian is calculated up to
high orders (1-triplon terms: 11$^{\rm th}$, 2-triplon terms:
10$^{\rm th}$ order). The effective observable $\mathcal{O}_{\rm
eff}$ is computed to order $10$ in the 2-triplon sector.

The resulting plain series are represented in terms of the
variable $1-\Delta^{\rm SG}/(J_\parallel+J_\perp)$ \cite{schmi03a,schmi03d}
where $\Delta^{\rm SG}$ is the one-triplon gap. Then
standard Pad\'e extrapolants \cite{domb89} yield reliable results up to
$J_\parallel/J_{\perp}= 1-1.5$ depending on the value of $J_{\rm
  cyc}/J_{\perp}$. Consistency checks were carried out by extrapolating the
involved quantities before and after Fourier transforms. In case of
inconclusive extrapolants the bare truncated series are used. We will estimate
the overall accuracy below by comparing with DMRG results\cite{nunne02}. The Raman line shape is
finally calculated as continued fraction by tridiagonalization of the effective two-triplon Hamiltonian.

Sectors with odd number of triplons are inaccessible by Raman
scattering due to the invariance of the two observables
$\mathcal{O}^{\rm leg}_{\rm eff}$ and $\mathcal{O}^{\rm
rung}_{\rm eff}$ with respect to reflections about the centerline
of the ladder\cite{schmi01}. Thus only excitations with even
number of triplons matter. Therefore the leading contributions to
the Raman response come from the 2-triplon sector. It was shown
earlier that the two-triplon contribution is the dominant part of
the Raman response at low and intermediate
energies\cite{schmi01,Diss_ks}. The role of the four-triplon
contribution for the high-energy spectral weight will be discussed
at the end of this work.

\section{Cuprate Ladders}

\label{cl}

In this part we will compare the theoretically obtained two-triplon
contributions to the experimental line-shapes of the cuprate ladders
Sr123 and Sr14.
 The crystals of Sr123 have been grown and measured under the same
conditions as described in Refs.~\onlinecite{loeff02} and
\onlinecite{goess03}, while the data of Sr14 have been provided by
Gozar {\it et al.}\cite{gozar01}. The experimental Raman
line-shape depends strongly on the laser energy because
resonant contributions are present. This becomes apparent in a
strong anisotropy between the width of the two-triplon peak in leg and rung
polarization for laser energies $\omega_{\mathrm{exc}}= 2-3$ eV. The width of the two-triplon peak in leg
polarization is much sharper. For laser energies
$\omega_{\mathrm{exc}}<2$ eV, the strong anisotropy between both polarizations vanishes\cite{gozar01}. It is
therefore important to figure out which laser energy has to be
used for the comparison between the non-resonant theory and the experiment in order to study the magnetic excitations only.

The first criterion can be gained from the optical conductivity as
for example given in Ref. \onlinecite{gozar01}: the intensity of
the two-triplon peak develops in the same way as the optical
conductivity. For the non-resonant regime both, energy of the
incident and scattered light, should be smaller than the charge
transfer gap (Sr14: $\Delta_{T={\rm 10 K}}\sim2.1$ eV).
Thus, we have chosen spectra with laser energies $\omega_{\mathrm{exc}}=1.92$
eV in the case of Sr14 and $\omega_{\mathrm{exc}}=1.95$
eV for Sr123. Luckily, the value of the optical
conductivity is about 100 $\Omega^{-1}$cm$^{-1}$ in the sub
gap regime at $\omega=\omega_{\mathrm{exc}}-E_{\rm 2T}\sim 1.5$ eV\cite{gozar01}
(which is 1-2 orders of magnitude larger than for the 2D
cuprates \cite{choi99}) yielding a non vanishing intensity of
the two-triplon peak. Here $E_{\rm 2T}$ denotes the energy of the two-triplon peak.

The second criterion arises from the polarization dependence of
the two-triplon peak. Depending on the laser energy used one can
observe a drastic difference in the line shape between the two
polarizations\cite{gozar01,goess03}. While the
difference in the line shapes is large for $\omega_{\mathrm{exc}}>2.1$
eV, it does almost vanish in the case of $\omega_{\mathrm{exc}}<2.0$ eV
\cite{gozar01}. This fits very well to the weak polarization dependence of the purely magnetic response
as described by Eqs.~(\ref{Observables}): for $x_{cyc}=0.0$ the Raman line-shape is identical in the
 rung and the leg polarization. Small deviations $x_{cyc}=0.2$ as relevant for the description of
Sr123 and Sr14 produce small
deviations with respect to the symmetry between the rung and leg
polarization of the ladder. These deviations cannot account for the
drastic change between the two polarizations as observed for
$\omega_{\mathrm{exc}}>2.1$ eV \cite{gozar01,goess03}. We therefore
conclude that the spectra $\omega_{\mathrm{exc}}<2.0$ eV are the best
choice in order to compare to a purely magnetic, non-resonant theory.

\begin{figure}[htbp]
  \begin{center}
    \includegraphics[width=8.2cm]{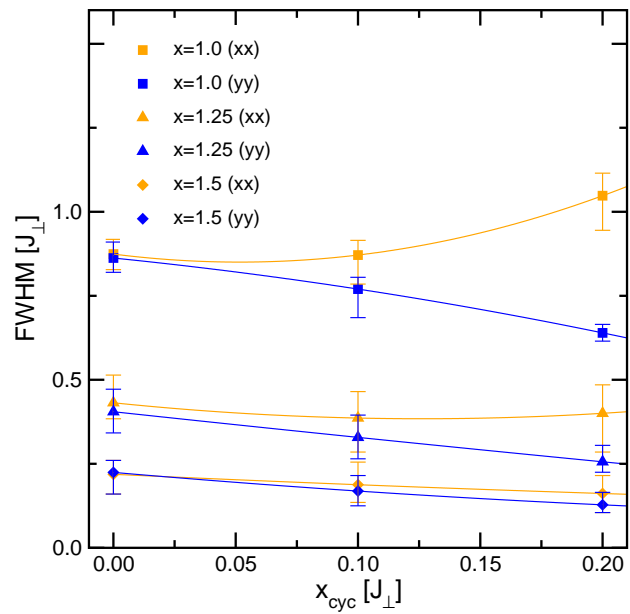}
    \caption{{\it Color online} The FWHM of the two-triplon peak for $x:=J_{\parallel}/J_{\perp}=1$ (squares), $x=1.25$ (triangles) and
$x=1.5$ (diamonds) as a function of the strength of the four-spin
interactions $x_{\rm cyc}:=J_{\rm cyc}/J_{\perp}$. The Orange (Grey)
symbols denote (xx)-polarization and the blue (black) symbols
(yy)-polarization. The solid lines are a guide to the eye
obtained by spline interpolation.}
    \label{fig:FWHM}
  \end{center}
\end{figure}

Now we discuss the dependence of the width of the two-triplon peak on the parameters $x$ and $x_{\rm
  cyc}$. In Fig.~\ref{fig:FWHM}, the full width at half maximum (FWHM) of the
two-triplon peak is shown. The overall uncertainty shown as error
bars in Fig.~\ref{fig:FWHM} of the extrapolated
two-triplon FWHM was determined by comparing to DMRG data\cite{error}.

Let us first consider the case $x_{\rm cyc}=0.0$. Here the
two-triplon width should be identical in both polarizations. It
can be clearly seen that the numerically obtained results reflect
this property rather well indicating that the uncertainties in
the extrapolation are small concerning the matrix elements. There
is a strong dependence of the FWHM of the two-triplon peak on the
parameter $x$. The peak sharpens significantly when the ratio $x$ of the magnetic Heisenberg exchanges
increases (4 times from $x=1$ to $x=1.5$).
In the case of $x_{\rm cyc}\neq 0$, the width depends on the polarization. In general, the width in
(xx)-polarization is larger than in (yy)-polarization. For fixed $x$ the FWHM changes at
maximum by a factor of two when varying $x_{\rm cyc}$ from $0$ to $0.2$.

\begin{figure}[htbp]
  \begin{center}
    \includegraphics[width=8.2cm]{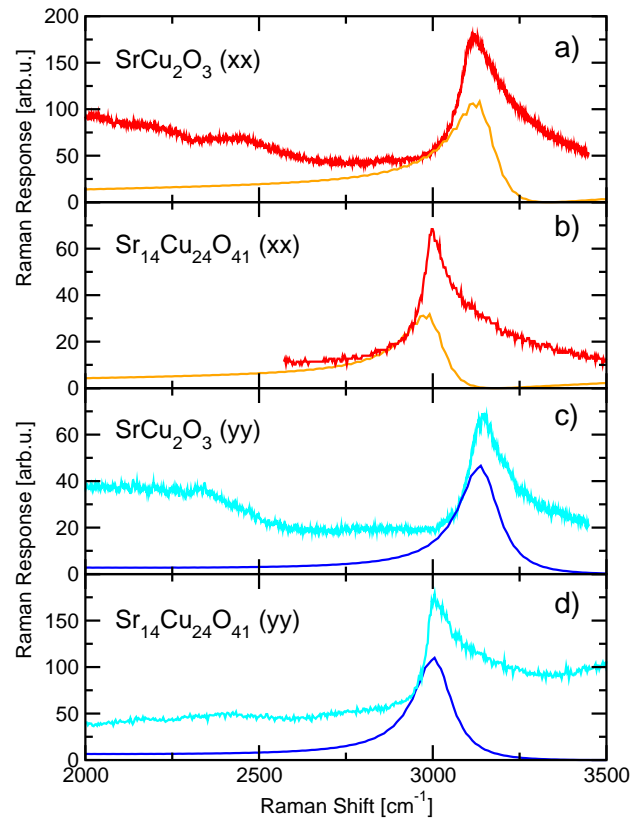}
    \caption{{\it Color online} Comparison of the magnetic Raman response of Sr123 ($T=25$ K) and
Sr14 ($T=5$ K) with the theoretically obtained two-triplon
contribution. The data of Sr14 have been provided by Gozar {\it
et al.}\cite{gozar01}. (a) The red (black) curve denotes the (xx)-polarization
($x\parallel b$) of Sr123
 with a laser excitation energy $\omega_{\mathrm{exc}}=$1.95 eV. The orange (grey)
curve displays the theoretical two-triplon contribution with $x=1.5$,
$x_{\rm cyc}=0.2$ and $J_{\perp}=1130$ cm$^{-1}$. (b) The red (black)
curve denotes the (xx)-polarization ($x\parallel a$) of Sr14 with
a laser excitation energy $\omega_{\mathrm{exc}}=1.92$ eV. The orange (grey) curve displays the
theoretical two-triplon contribution with $x=1.5$, $x_{\rm
cyc}=0.2$ and $J_{\perp}=1080$ cm$^{-1}$. (c) (yy)-polarization
($y\parallel a$) for Sr123. Identical parameters as in (a). The cyan (grey)
curve displays the experimental data and the blue (black) curve the theoretical
two-triplon contribution. (d) (yy)-polarization ($y\parallel c$) for Sr14. Identical
parameters as in (b) and the same colors as in (c).}
    \label{fig:Comp}
  \end{center}
\end{figure}

In Figs.~\ref{fig:Comp}(a-d), the experimental Raman response of
Sr123 and Sr14 is shown for (xx)- and (yy)-polarization (red/black and cyan/grey curves).
The spectra of Sr123 were taken in the same
way as described in Ref. \onlinecite{goess03}. The data of Sr14
has been made available by Gozar {\it et al.}\cite{gozar01}. In
addition, theoretically obtained two-triplon contributions are
displayed (orange/grey and blue/black). Experimentally, the width of
the two-triplon peak of both materials is almost identical ($\sim$
150 cm$^{-1}$). Only the position of the two-triplon peak is
different ($\sim$ 3140 cm$^{-1}$ for Sr123 and $\sim$ 3000
cm$^{-1}$ for Sr14) which is a result of the slightly different
Madelung potentials of both compounds.

For modeling the Raman response we assume $x_{\rm cyc}=0.2$ for both
compounds. This order of magnitude was previously obtained for cuprate ladders by inelastic neutron
scattering\cite{matsu00a,matsu00b}, by infrared
absoption\cite{nunne02,gruni03}, Raman response\cite{schmi03b,goess03}
and theoretical works deriving extended low-energy Heisenberg models\cite{mizun97,calza03}.
In order to account for the FWHM and the two-triplon peak position we determine $x=1.5$ and global energy scales $J_{\perp}=
1130$ cm$^{-1}$ for Sr123 and $J_{\perp}=1080$ cm$^{-1}$ for Sr14.
It was previously argued that in Sr14 a charge order of the chain
subsystem modulates the magnetic exchange in the ladders\cite{schmi03b}. This opens a gap in the Raman
response which has a large effect on the two-triplon peak for
the parameters $x=1.2$ and $x_{cyc}=0.2$ which are appropriate
for La6Ca8. However, the effect is small for larger $x$-values because the induced
gap opens well above the two-triplon peak at $\sim 3600$ cm$^{-1}$.
The set of parameters used above for Sr123 and Sr14 describes quantitatively well the Raman response as shown in
 Tab.~\ref{tab_plane} and Figs.~\ref{fig:Comp}(a-d). Especially both polarizations for each material can be
 modeled using only one set of parameters $J_\perp$, $x$ and $x_{\rm cyc}$. The smaller FWHM of Sr123 and Sr14 compared to
La6Ca8\cite{sugai99,schmi03b} can be directly explained by their
larger $x$-values (see Fig.~\ref{fig:FWHM}). The coupling
constants of Sr14 are in good agreement with those obtained by IR
absorption measurements\cite{windt03}. Additionally, our set of parameters
yields a spin gap of $290$ cm$^{-1}$ for Sr123 and $280$ cm$^{-1}$for Sr14 using
the underlying one-triplon dispersion. The latter value is consistent with the spin gap measured by inelastic neutron scattering\cite{eccle98}.

\section{Cuprate Planes}
\label{cp}

\begin{figure}[htbp]
    \begin{center}
     \includegraphics[width=\columnwidth]{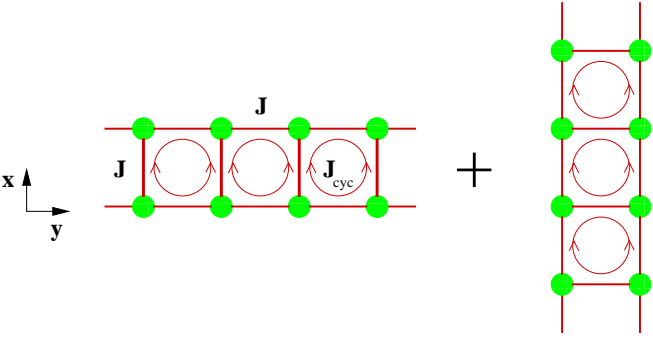}
    \end{center}
    \caption{Sketch of two uncoupled spin-ladders. Here one ladder is oriented in
      $y$-direction and the other in $x$-direction. We approximate
      the two-dimensional square lattice by the sum of these two uncoupled
      orthogonal ladders.}
    \label{fig_sketch_plane}
\end{figure}

In this section we calculate the Raman response for the undoped
two-dimensional cuprate compounds using a toy model consisting of
two uncoupled two-leg ladders (see Fig.~\ref{fig_sketch_plane}).
This is motivated by the fact that the building blocks of ladders
and planes are edge-sharing Cu$_4$ plaquettes. We expect that the
Raman response is dominated by short-range and high-energy
excitations yielding a certain similarity between ladders and
planes. Indeed, the positions of the two-magnon peak in the 2D
cuprates and the two-triplon peak in the cuprate ladders are
found at almost the same frequency $\sim 3000$ cm$^{-1}$, but the
FWHM of the two-dimensional compounds is a factor of 2-6 larger.
We have shown in the last section that the FWHM of the
two-triplon peak in the cuprate ladder compounds strongly varies
with $x$. We therefore conjecture that the larger FWHM of the
two-dimensional cuprates originates from the isotropic coupling
$x=1$. There will be of course deviations at small energies
resulting from the differences between a gapped two-leg ladder
and the gapless excitations in the two-dimensional compounds.
Clearly, a magnon description would be the proper starting point
to treat the long-ranged ordered antiferromagnetic state. We
think, however, that a triplon picture which includes the
interactions on the quantitative level can give a good
description of the Raman response. A similar treatment in terms
of gapped quasi-particles already led to an improved agreement
between theory and experiments\cite{hsu90,wang91}.

In the following we will show how to deduce the A$_{\rm
1g}$ and B$_{\rm 1g}$ Raman spectra of the square lattice from those of the two-leg
  ladder. Clearly, one should use $x=1$ because the square lattice is
  isotropic ($J=J_\parallel=J_\perp$). Starting from the Fleury-Loudon operator
 the observables $\mathcal{O}^{\rm B1g}$ ($\mathcal{O}^{\rm A1g}$) for magnetic
light scattering in B$_{\rm 1g}$ (A$_{\rm 1g}$) polarization read in leading order for the
two-dimensional square lattice\cite{Fleury1,Shastry1}
\begin{subequations}
\label{Observables_2D}
\begin{eqnarray}
 \mathcal{O}^{\rm B1g}& =& A_{\rm 0,B1g} \left( \sum_{<ij>,x}
 {\bf S}_{i}{\bf S}_{j} - \sum_{<ij>,y} {\bf S}_{i}{\bf S}_{j} \right)\\
 \mathcal{O}^{\rm A1g}& =&  A_{\rm 0,A1g} \left( \sum_{<ij>,x}
 {\bf S}_{i}{\bf S}_{j} + \sum_{<ij>,y} {\bf S}_{i}{\bf S}_{j} \right)\ .
\end{eqnarray}
\end{subequations}

Here $<ij>,x$ ($<ij>,y$) denotes a summation over
nearest-neighbors in $x$-direction ($y$-direction). The parameters $A_{\rm
  0,B1g}$ and $A_{\rm 0,A1g}$ depend on the underlying microscopic model and
are in general not equal\cite{Shastry1}. We approximate the
two-dimensional square lattice by a sum of two
uncoupled two-leg ladders, one oriented in $x$-direction, the
other in $y$-direction. The situation is sketched in
Fig.~\ref{fig_sketch_plane}. The summation over both ladder
orientations will restore the square lattice symmetries.
Comparing Eq.\ (\ref{Observables_2D}) with Eq.\
(\ref{Observables}) one readily deduces the following relations

\begin{subequations}
\label{Observables_Rel}
\begin{eqnarray}
 \mathcal{O}^{\rm B1g}& \propto& \left( \mathcal{O}^{\rm leg}-\mathcal{O}^{\rm
 rung}\right) \\
 \mathcal{O}^{\rm A1g}& \propto& \left( \mathcal{O}^{\rm leg}+\mathcal{O}^{\rm rung}\right)
\end{eqnarray}
\end{subequations}
between the relevant observables in the two-leg ladder and the
two-dimensional square lattice. Note that for $x_{\rm cyc}=0$,
the Raman response in the A$_{\rm 1g}$ polarization vanishes due
to the property $\mathcal{O}^{\rm leg}|0\rangle =
-\mathcal{O}^{\rm rung}|0\rangle$\cite{schmi01}. The latter point
is consistent with earlier treatments of the two-dimensional
Raman response. But for a finite strength of the four-spin
interactions $x_{\rm cyc}$, also the A$_{\rm 1g}$ polarization is
finite\cite{honda93,erole99}.

\begin{figure}[htbp]
  \begin{center}
    \includegraphics[width=8.2cm]{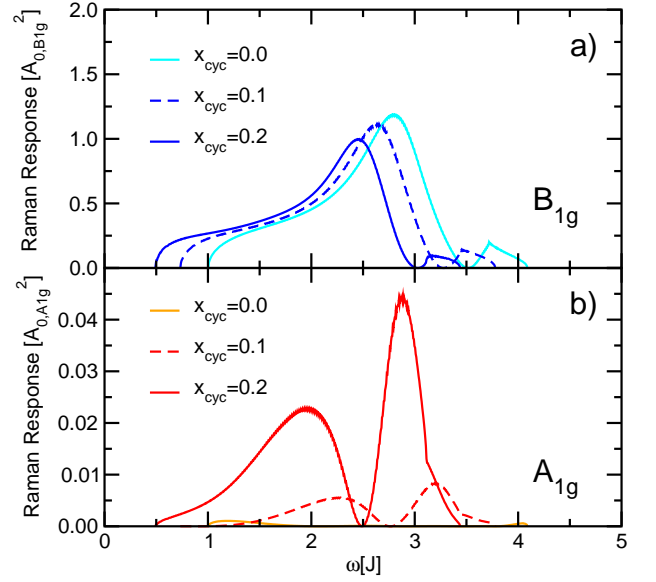}
    \caption{{\it Color online} Two-triplon Raman response of the 2D square lattice
      for $x=1$. (a) B$_{\rm 1g}$-polarization for
      $x_{\rm cyc}=0.0$ (cyan/grey), $x_{\rm cyc}=0.1$ (dashed) and
      $x_{\rm cyc}=0.2$ (blue/black). (b) A$_{\rm 1g}$-polarization  for
      $x_{\rm cyc}=0.0$ (orange/grey), $x_{\rm cyc}=0.1$ (dashed) and
      $x_{\rm cyc}=0.2$ (red/black). Note the different scales for the Raman
      response in $A_{\rm 1g}$ and $B_{\rm 1g}$ spectra.}
    \label{fig:2d_A1g_B1g}
  \end{center}
\end{figure}

The theoretical two-triplon contribution of the B$_{\rm 1g}$ (panel a) and the A$_{\rm
  1g}$ (panel b) polarization is shown in
  Fig.~\ref{fig:2d_A1g_B1g}. The parameters used are an isotropic coupling $x=1$
  and a strength of the four-spin interactions $x_{\rm cyc}=0.0$,
  $x_{\rm cyc}=0.1$, and $x_{\rm cyc}=0.2$.

The B$_{\rm 1g}$ polarization displays a symmetric two-triplon
peak which is dominating the Raman response. The four-spin
interactions shift the whole spectrum to lower energies and
decrease the total intensity. The FWHM of the two-triplon peak is
approximately given by the average width of the two-triplon peaks
of an isolated two-leg ladder in rung and in leg polarization. Thus, the width
is nearly independent of the value of $x_{\rm cyc}$.

The A$_{\rm 1g}$ polarization is almost zero for vanishing
$x_{\rm cyc}$. This again reflects the accurate extrapolation of
the matrix elements. For non-zero $x_{\rm cyc}$, a finite A$_{\rm
1g}$ contribution is realized. The differences in the line-shape
between A$_{\rm 1g}$ and B$_{\rm 1g}$ are a pure effect of
different matrix elements. Compared to the two-triplon peak in the
B$_{\rm 1g}$ polarization the A$_{\rm 1g}$ polarization displays a
two peak structure where the second peak is sharper and at higher
energies.

\begin{table*}
\caption{\label{tab_plane}Comparison of the two-triplon peak between experimental data of different
  cuprate ladder and plane compounds and the theoretical results.}
\begin{ruledtabular}
\begin{tabular}{cccccccccc}
 &\multicolumn{5}{c}{Experiment}&\multicolumn{4}{c}{Theory (CUT)}\\
 Material&Peak&FWHM$^a$&$\omega_{\mathrm{exc}}$&Ref.&$x$&$x_{cyc}$&$J_\perp$&FWHM\\
 &[cm$^{-1}$]&[cm$^{-1}$]&[eV]& & & &[cm$^{-1}$]&[cm$^{-1}$]\\
 \hline
  SrCu$_2$O$_3$ (xx)        & 3120 & 150-220 & 1.96 & this work & 1.5 & 0.2 & 1130 & 180 \\
  SrCu$_2$O$_3$ (yy)        & 3150 & 120-180 & 1.96  & this work & 1.5 & 0.2 & 1130 & 140 \\
  Sr$_{14}$Cu$_{24}$O$_{41}$ (xx)& 3000 & 100-160 & 1.92  & \onlinecite{gozar01} & 1.5 & 0.2 & 1080 & 180 \\
  Sr$_{14}$Cu$_{24}$O$_{41}$ (yy)& 3000 & 120$^b$ & 1.92  & \onlinecite{gozar01} & 1.5 & 0.2 & 1080 & 140 \\
  La$_{6}$Ca$_{8}$Cu$_{24}$O$_{41}$ (xx)& 3010 & 550$^b$ & 2.41  & \onlinecite{sugai99,schmi03b} & 1.2 & 0.2 & 1130 & 580 \\
  La$_{6}$Ca$_{8}$Cu$_{24}$O$_{41}$ (yy)& 2950 & 350$^b$ & 2.41  & \onlinecite{sugai99,schmi03b} & 1.2 & 0.2 & 1130 & 350 \\
 \hline
  Sr$_2$CuO$_2$Cl$_2$ & 2950 & 800-1100 & 2.73  & \onlinecite{blumb96} & 1.0 & 0.2 & 1190 & 1000  \\
  YBa$_2$Cu$_3$O$_{6+\delta}$ & 2750 & 1000-1150 & 2.71  & \onlinecite{sugai90} & 1.0 & 0.2 & 1110 & 940 \\
  La$_2$CuO$_4$ & 3170 & 950-1150 & 2.71  & \onlinecite{sugai90} & 1.0 & 0.2 & 1280 & 1080 \\
  Nd$_2$CuO$_4$ & 2930 & 900-1050 & 2.71  & \onlinecite{sugai90} & 1.0 & 0.2 & 1190 & 1000 \\
\end{tabular}
\end{ruledtabular}
\begin{tabular}{l}
$^a$ Lower limit of exp. FWHM: linear background subtracted from
data. Upper limit: no background corrections. \\
$^b$ Exp. FWHM: linear background subtracted from data because
background exceeds almost the two-triplon peak heights.
\end{tabular}
\end{table*}

In the following we will compare the theoretical two-triplon
contribution to the Raman response with low temperature
experimental data on R$_2$CuO$_4$ ($\omega_{\mathrm{exc}}=2.71$
eV)\cite{sugai90},
  Sr$_2$CuO$_2$Cl$_2$ ($\omega_{\mathrm{exc}}=2.73$ eV)\cite{blumb96}, and
YBa$_2$Cu$_3$O$_{6+\delta}$ ($\omega_{\mathrm{exc}}=2.71$
eV)\cite{sugai90} taken from the literature.

As discussed in section \ref{cl} the laser energy used for the
experiment is a crucial issue. Analogous to the ladders one should
use spectra of cuprate planes measured with laser energies
below the charge gap for comparing to the purely magnetic
theoretical response. But it turns out that the optical
conductivity is rather low ($\lesssim10$ $\Omega^{-1}$cm$^{-1}$)
below the charge gap $\Delta$ which results in a vanishing
intensity of the two-magnon peak \cite{blumb96}. An analogous choice of the
laser energies below the charge gap as discussed for the cuprate ladders is
not possible.
Therefore, we used data measured with laser energies $\omega_{\mathrm{exc}}=2.7$ eV$>\Delta$. At the energy $\omega_{\mathrm{exc}}=2.7$ eV
the optical conductivity is quite smooth and $\omega_{\mathrm{exc}}\neq\Delta\approx (1.7-2.0)$ eV\cite{salam95,blumb96,choi99}.
Simultaneously $\omega_{\mathrm{exc}}=2.7$ eV coincides with the
triple resonance at $\omega_{\rm res}\approx \Delta+8J$. The triple resonance theory predicts
two peaks in the Raman response at about $2.8 J$ and $4 J$. The
relative intensity of both peaks depends on the laser energy. The second peak
is strongly suppressed at $\omega_{\rm res}$. In that sense this laser energy
can be assigned to be closest to the non-resonant regime \cite{chubu95,schoe97,morr97}.

\begin{figure}[htbp]
  \begin{center}
    \includegraphics[width=8.2cm]{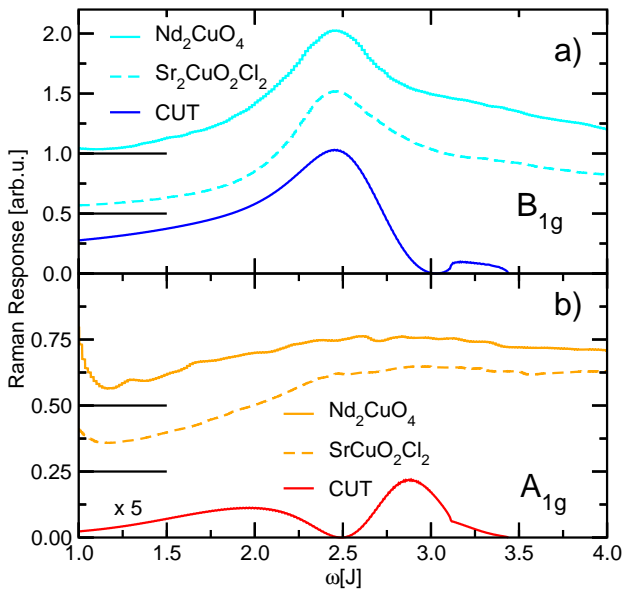}
    \caption{{\it Color online} Comparison of the two-triplon Raman response to
      the two-magnon Raman line-shape of Nd$_2$CuO$_4$
      ($\omega_{\mathrm{exc}}=2.71$ eV, $T=30$ K) and
  Sr$_2$CuO$_2$Cl$_2$ ($\omega_{\mathrm{exc}}=2.73$ eV, $T=5$ K). The
  Raman data of Nd$_2$CuO$_4$ and Sr$_2$CuO$_2$Cl$_2$ are reproduced from
  Refs.~\onlinecite{sugai90} and \onlinecite{blumb96}. The experimental
  curves are smoothed and their zero position is shifted horizontally as indicated by the black horizontal
  lines. (a) B$_{\rm 1g}$-polarization: The blue (black) curve denotes the two-triplon
  contribution with $x=1$ and $x_{\rm cyc}=0.2$. The global energy scale $J$ is
  chosen such that experimental two-magnon and the theoretical two-triplon
      peak merge. This yields $J=1190$ cm$^{-1}$ for Nd$_2$CuO$_4$ (cyan/grey) and
  Sr$_2$CuO$_2$Cl$_2$ (dashed). (b) A$_{\rm 1g}$-polarization (theory:
      red/black; experiment: organge/grey): Same notations as
  in (a). Note that the {\it same} magnetic exchange couplings $J$ are
      used. The A$_{\rm 1g}$-CUT is multiplied by a factor 5 in comparison to
      the curve in Fig.~\ref{fig:2d_A1g_B1g}.}
    \label{fig:2d_Comp}
  \end{center}
\end{figure}

In Fig.~\ref{fig:2d_Comp} experimental data and theoretical
contributions using $x=1$ and $x_{\rm cyc}=0.2$ are shown for both
polarizations. Frequencies are measured in units of $J$. We first discuss
the B$_{\rm 1g}$ polarization in Fig.~\ref{fig:2d_Comp}(a). We
have chosen the global energy scale $J$ for all experimental
curves such that the positions of the experimental two-magnon and the theoretical
two-triplon peaks match. This yields quantitatively reasonable
values for these compounds. In addition, we find quantitative
agreement between the experimental FWHM and the theoretical FWHM of
the two-triplon peak. The values of $J$ and the FWHM are listed in
Tab.~\ref{tab_plane} for all compounds. Note that the FWHM for $x=1$ is larger
than for the anisotropic case $x>1$ as discussed for the ladder compounds.

 Clearly, there are also deviations
  between theory and experiment. As expected, the low-energy spectral weight
  in the theoretical line-shape is larger compared to the experimental
  curves. This is definitely a consequence of approximating the
  two-dimensional square lattice with quasi one-dimensional models. There is
  also spectral weight missing at higher energies above the two-triplon
  peak. Possible explanations will be described below.

The results for the A$_{\rm 1g}$ polarization (shown in Fig.
\ref{fig:2d_Comp}(b)) are explained next. We used the {\it same}
global energy scales $J$ for the experimental curves as determined
from the B$_{\rm 1g}$ polarization above. In order to reproduce
the maximum intensity of the experiment, we multiplied the
theoretical curve from Fig.~\ref{fig:2d_A1g_B1g}(b) by a factor
5. This implies that the microscopic parameters $A_{\rm 0,B1g}$
and $A_{\rm 0,A1g}$ are anisotropic. A possible reason for this
anisotropy could be the restriction to the Fleury-Loudon
observable. An extension of this observable to higher orders in
$t/U$ (four-spin and next-nearest neighbor two-spin terms) gives
additional contributions to $\mathcal{O}^{\rm
  B1g}$ and $\mathcal{O}^{\rm A1g}$\cite{Shastry1}. The relevance of these
contributions has not been analyzed.

In the experiment a broad hump is measured. We find it very
promising that the theoretical contribution displays the dominant
spectral weight just for these energies. However, the line-shape
can not be resolved completely because the dip in the theoretical
curve is not observed in the experiment. It originates from
neglecting the finite life-time effects which are already present
in the description of the isolated two-leg
ladder\cite{schmi03b,Diss_ks} being the building block of our
square-lattice toy model. We conclude that at least a part of the
experimental A$_{\rm 1g}$ polarization originates from the finite
four-spin interactions. For $x_{\rm cyc}=0$, there is no purely
magnetic contribution to the Raman response for this
polarization. A finite A$_{\rm 1g}$ Raman response can be
regarded as an evidence for the presence of sizable four-spin
interactions. This follows entirely from symmetry arguments and
holds true for the full two-dimensional model. At higher
energies, spectral weight is missing in the theoretical
contribution in an analogous fashion as in the B$_{\rm 1g}$
polarization.

\section{High-energy spectral weight}

\label{sw}

As shown in Sect.~\ref{cl} and in Sect.~\ref{cp} the CUT cannot
account for the missing high-energy spectral weight when
comparing to the Raman experiments. Also other theories proposed
previously like calculations based on
spin-waves\cite{singh89,canal92}, paramagnons\cite{hsu90},
Jordan-Wigner fermions\cite{wang91} and numerical
studies\cite{sandv98,Suzuki1,Jurecka1,schmi01} were faced with
the same problem. Extended theories including {\it (i)
multi-particle contributions}, {\it (ii)
  spin-phonon coupling}, {\it (iii) two-magnon/triplon plus phonon absorption}
and {\it (iv) triple resonance} are necessary in order to describe the high-energy spectral weight.

Most of the publications deal with the two-dimensional compounds. Here we will try to
review these ideas and reexamine them in the light of our new
results. Especially the quantitative results for the cuprate
ladders can give new insights in this discussion.

{\it (i) multi-particle contributions} One open problem is the
role of multi-particle contributions to the Raman response, i.e.
the four-magnon contribution in the case of the square lattice
and the four-triplon contribution in the case of the two-leg
ladder. At this stage no quantitative calculations are available.
But it is known that the multi-triplon spectral weights are
sizable for the two-leg ladder\cite{schmi01,Diss_ks}. The main
effect of the four-spin interaction on the high-energy spectral
weight is a small shift from the two-triplon to the multi-triplon
channels\cite{Diss_ks}. But this shift is not sufficient to
account for the high-energy spectral weight as observed in
experiments. This was also found in treatments for the
two-dimensional square lattice\cite{honda93,loren99,katan03}.
However, the complete magnetic infrared absorption spectrum
(including the high-energy part) of La6Ca8 can be described
quantitatively by including multi-particle
contributions\cite{nunne02,gruni03}. Here $x_{\rm cyc}$ does not
play the dominant role for the high-energy spectral
weight\cite{windt01}.

It is therefore plausible that these contributions give a
noticeable effect also on the high-energy Raman response.
There are also indications that the spectral weight cannot be
fully explained in this way. For example, the four-magnon spectral weight was
shown to be negligible for the 2D square lattice\cite{canal92}. But the magnon-magnon
interaction which was not treated in this calculation could enhance the
high-energy spectral weight. Also quantum Monte Carlo
calculations which include all magnon contributions for the
two-dimensional Heisenberg model seem to explain only a part of the
high-energy spectral weight\cite{sandv98}. But finite size effects
and inaccuracies of the analytical continuation can lead to uncertainties in
determining the high-energy spectral weight.

{\it (ii)  spin-phonon coupling} The latter observations suggest that additional degrees of
freedom are important. It was argued by several authors that the
coupling to phonons produces a large amount of spectral weight
above the two-triplon
peak\cite{erole99,saeng95,weber89,haas94,nori95,knoll90}. In one approach
the spin-phonon coupling modulates the magnetic exchange couplings with a
Gaussian distribution. Another approach introduces a finite spin wave damping
induced by the spin-phonon coupling. Both scenarios produce a
significantly broadened and asymmetric two-magnon peak as
observed in experiments\cite{haas94,nori95}.

Nevertheless, the consistency of a spin-phonon coupling as
suggested above with experiments is not clear. The magnitude of
this coupling has to be unrealistically large in order to
describe infrared absorption data\cite{gruni00}. Additionally, it
was pointed out by Freitas and Singh\cite{freit00} that the
temperature-dependent correlation length and the spin dynamics
which agree well with purely magnetic models does not leave room
for such a coupling\cite{endoh88,chakr89}.

There are no investigations of the role of spin-phonon couplings
for the case of the cuprate ladder systems. But the FWHM of the two-triplon
peak can be quantitatively understood within a purely magnetic model as shown
in Sect.~\ref{cl}. Thus, we conclude that the spin-phonon coupling is not strong in the case of the cuprate ladder compounds. Such a
coupling leads to a broad two-triplon peak in the same way as for
the two-dimensional case. This is a contradiction when
considering the Raman response and the infrared absorption of
cuprate ladders simultaneously: on the one hand one needs a larger
anisotropy between leg and rung coupling (larger $x$) in order to
sharpen the two-triplon peak in the Raman response again (see Fig.~\ref{fig:FWHM}) but on
the other hand one cannot explain the infrared absorption with
an substantially increased $x$\cite{nunne02,windt03}. A strong spin-phonon coupling is
therefore in contradiction with the results obtained for cuprate
ladders. This can be also seen as an indication that the same holds true
for the two-dimensional compounds\cite{knoll90}.

{\it (iii) two-magnon/triplon plus phonon absorption} A third
alternative explaining the high-energy spectral weight uses
phonons as possible momentum sinks. Here a strong spin-phonon
coupling is not necessary. The idea is based on the work of
Lorenzana and Sawatzky for infrared
absorption\cite{loren95a,loren95b,loren99}. It is well accepted
in the case of infrared absorption measurements on cuprate
ladders\cite{windt01,nunne02} and planes\cite{loren99,gruni00}
that the dominant processes are magnetic excitations which are
assisted by phonons. It was realized by Freitas and Singh that
similar processes could be important also for the Raman response
in cuprate planes\cite{freit00}.  In an analogous fashion a
two-triplon plus (Raman active) phonon process for the Raman
response in cuprate ladders could be important. It can be used to
transfer spectral weight above the two-triplon peak leading to an
asymmetric line-shape. It is a difficult task to determine the
relative strength of this process compared to the usual
two-triplon scattering.

{\it (iv) triple resonance} Additionally, the triple resonance was proposed to account for
the high-energy spectral weight in the two-dimensional
compounds\cite{chubu95,schoe97,morr97,blumb96}. As already stated
in Sec.~\ref{cp}, the experimental spectra of the planes are
taken in the resonant regime. It is known that the triple
resonance scenario yields an additional peak above the two-magnon
peak. Its intensity depends significantly on the energy of the
incident light in accordance with experiments\cite{blumb96}. In
principle, the same effect is also present in ladder compounds.
But for the laser energy $\omega_{\mathrm{exc}}$=1.92 eV$<\Delta$ considered for Sr123 and Sr14 the triple resonance condition is not fulfilled.

Due to the simplified model used for the 2D system no conclusion about
the high-energy weight can be drawn from our results. Because a large spin-phonon coupling and the triple resonance can be ruled
out for the cuprate ladder systems, the observed high-energy spectral weight in the cuprate ladder compounds has to
be explained most probably by the multi-triplon or two-triplon plus phonon contributions.

\section{Conclusion}

\label{conclusion}

The first part of this work deals with the theoretical understanding of
non-resonant magnetic Raman scattering on cuprate two-leg ladder
compounds, namely Sr123, Sr14, and La6Ca8. Therefore we applied a
triplon-conserving CUT on a microscopic spin-model which includes
Heisenberg couplings and additional four-spin interactions. We
studied the two-triplon contribution to the non-resonant magnetic
Raman response. The dominating feature of the two-triplon
contribution is the two-triplon peak which has a characteristic
FWHM depending on the model parameters $x$ and $x_{\rm cyc}$.
We carefully chose the experimental data closest to
the non-resonant regime and compared them with our theory.

The key observation we found is that the sharpness of the
two-triplon peak in Sr123 and Sr14 in comparison to La6Ca8 can be explained by the
stronger anisotropy of the magnetic exchange along the rungs and legs of
the ladder. Indeed, the two-triplon peak width depends strongly on
the parameter $x$. Both materials can be modeled with the
parameters $x\approx 1.5$ and $x_{\rm
  cyc}\approx 0.2$ but different global energy scales $J_{\perp}\approx 1130$
cm$^{-1}$ for Sr123 and $J_{\perp}\approx 1080$ cm$^{-1}$ for
Sr14. The parameters for Sr14 are in good agreement with infrared
absorption \cite{nunne02,windt03} and inelastic neutron
scattering\cite{eccle98} experiments.
We conclude that the dominating two-triplon peak of the magnetic Raman response in cuprate
ladders can be consistently explained within the microscopic
model. The presence of a four-spin interaction of the order of
$0.2 J_\perp$ can be viewed as a settled issue.

In the second part of this article we used the results found for
the two-leg ladder to describe the magnetic Raman response of the
undoped two-dimensional cuprate compounds in B$_{\rm 1g}$ and
A$_{\rm 1g}$ polarization. The contribution to the A$_{\rm 1g}$
polarization is only allowed for finite four-spin interactions
due to symmetry reasons. We use an isotropic coupling $x=1$ and
$x_{\rm cyc}=0.2$ for the comparison with the experimental data.
Convincingly, we find quantitative agreement for the two-triplon
peak position {\it and} the two-triplon peak width for several
compounds. Additionally, a sizable spectral weight is found in
the A$_{\rm 1g}$ polarization consistent with experiments. We
conclude that the processes dominating the magnetic Raman
response are short-ranged.

The last part deals with the missing high-energy
spectral weight above the dominating two-triplon peak for the
case of cuprate ladders and planes. We review possible sources of
this spectral weight like multi-particle contributions, the role
of spin-phonon coupling, a two-triplon plus phonon process and
the triple resonance to the magnetic Raman response. We deduced
from our results that the high-energy spectral weight cannot be explained with
realistic values for the spin-phonon coupling.

In summary, our calculations lead to an unified understanding of the magnetic
Raman response in cuprate ladder compounds within a purely magnetic model. A
strong spin-phonon coupling can be excluded for these materials. Additionally,
we obtained a convincing quantitative description of the dominating
two-magnon peak in the Raman response of cuprate
planes using a toy model consisting of two uncoupled two-leg ladders.
This suggests that the short-ranged triplon excitations might be an
alternative starting point for the description of the two-dimensional cuprate compounds.

\section*{Acknowledgment}

We gratefully acknowledge G.S. Uhrig, M.\ Gr\"uninger, A. Gozar, A. Reischl, and S. Jandl
for very helpful discussions. We thank A. Gozar
for the kind provision of experimental data on
Sr14. This work is supported by the DFG in SFB 608.
\vspace*{+5mm}


\begin{thebibliography}{10}
\bibitem{sr14a}
Sr14 has intrinsically six holes per unit cell, but
it is believed that the majority of the holes are located in the
chain subsystem (see Ref.~\onlinecite{nueck00}).
\bibitem{notiz1}
Neglecting the two-spin exchange terms along the diagonals,
our notation
$J_\perp, J_\parallel, J_{\rm cyc}$ is given in terms of the other
notation $J^{\rm p}_\perp, J^{\rm p}_\parallel, J^{\rm p}_{\rm
cyc}$ using cyclic permutations by $J_\parallel =J^{\rm
p}_\parallel +\frac{1}{4}J^{\rm p}_{\rm cyc}$, $J_\perp =J^{\rm
p}_\perp +\frac{1}{2}J^{\rm p}_{\rm cyc}$, and $J_{\rm cyc}=J^{\rm
p}_{\rm cyc}$ (see Ref.~\onlinecite{brehm99}).
\bibitem{error}
The extrapolation of the matrix elements are relatively
accurate. The maximum deviation between the two-triplon widths in
both polarizations is $0.03 J_\perp$ for $x_{\rm
  cyc}=0$. The one-triplon hopping elements yield a quantitative one-triplon dispersion when compared to
DMRG data (see Ref.~\onlinecite{nunne02}). So we conclude that the largest
extrapolation error is due to the extrapolation of the
two-triplon interaction amplitudes. To get an estimate of this
error we compared the energy of the $S=0$ two-triplon bound state
at finite momentum for $x_{\rm cyc}=0$ obtained by DMRG with the results of the CUT. Then we
changed the two-triplon interaction for the Raman case by the
maximum deviation of the bound state energies and calculated the
effect on the FWHM of the two-triplon peak. In the end we added
$0.03 J_\perp$ to account for the uncertainty of the matrix
elements. The resulting uncertainties are shown as error bars in
Fig.~ \ref{fig:FWHM}.
\end{thebibliography}
\end{document}